\documentstyle[preprint,aps,floats,epsfig]{revtex}
\draft
\tightenlines
\begin{document}
\title{
$J/\psi$ suppression in ultrarelativistic nuclear collisions
}
\bigskip
\author{
Bin Zhang$^{\rm a}$,
C. M. Ko$^{\rm a}$,
Bao-An Li$^{\rm b}$,
Ziwei Lin$^{\rm a}$,
and Ben-Hao Sa$^{\rm a}$
\footnote{Permanent address: China Institute of Atomic Energy, 
P.O. Box 275 (18), Beijing, 102413, P.R. China}}

\address{$^{\rm a}$Cyclotron Institute and Physics Department,\\
Texas A\&M University, College Station, TX 77843, USA}
\address{$^{\rm b}$Department of Chemistry and Physics\\
Arkansas State University, P.O. Box 419\\
State University, AR72467-0419, USA}
\maketitle

\begin{abstract}
Using a multiphase transport model, we study the relative 
importance of $J/\psi$ suppression mechanisms due to plasma screening, gluon 
scattering, and hadron absorption in heavy ion collisions at the
Relativistic Heavy Ion Collider. We find that for collisions
between heavy nuclei such as Au+Au, both plasma screening and gluon 
scattering are important. As a result, the effect due to 
absorption by hadrons becomes relatively minor. 
The final $J/\psi$ survival probability in these collisions
is only a few percent. In the case of collisions between light
nuclei such as S+S, the effect of plasma screening 
is, however, negligible in spite of the initial high parton density. 
The final $J/\psi$ survival probability thus remains appreciable after 
comparable absorption effects due to gluons and hadrons.
\end{abstract}

\section{introduction}

According to the fundamental theory of the strong interactions, 
Quantum Chromo-Dynamics (QCD), normal nuclear matter is expected
to undergo a phase transition to deconfined quarks and gluons
when its density and/or temperature are high \cite{bmuller1}.
To produce such a quark-gluon plasma (QGP) in the laboratory, experiments
involving collisions of nuclei at relativistic energies have
been carried out at the CERN SPS \cite{qm99a}.  Possible evidences for the
production of this new phase of matter have recently been announced
\cite{cern1}.  Further experiments at the Relativistic Heavy Ion Collider 
(RHIC), which allows collisions at much high energies than those available 
previously, are expected to provide a better opportunity to create the
quark-gluon plasma and to study its properties. 

Since quarks and gluons cannot be directly detected in experiments,  
many indirect observables have been proposed as possible signatures
for the QGP. These include: enhanced production of strange hadrons 
as a result of the short strangeness equilibration time in QGP 
\cite{Rafelski:1982pu}, suppression of $J/\psi$ production due to
color screening in QGP \cite{Matsui:1986dk}, quenching of high $p_t$ jets
due to passage through the QGP \cite{Wang:1992xy,Gyulassy:1994hr,bdmps1},
and enhancement of low mass dileptons as a result of medium modifications of
hadron properties in high density matter \cite{gli1,wcassing1}. 
Although all these signals have been observed in heavy ion collisions at 
CERN SPS, alternative explanations without invoking the formation of the
quark-gluon plasma have also been proposed. As the QGP is expected to be 
produced at RHIC, it is of interest to study 
these signatures in heavy ion collisions at such high energies. 
Since the quark-gluon plasma has a finite size, exists for a finite time,
and may not be in equilibrium, it is important to use a dynamical model
to take into account these effects.
Using the parton cascade model \cite{Geiger:1992nj}, Satz and
Srivastava \cite{Satz:2000wv} have recently studied the time evolution of 
the density profile of the parton system in these collisions
and demonstrated the possibility
of using a dynamical model to study the onset of
deconfinement and its effect on quarkonium suppression.
In this paper, we shall investigate explicitly $J/\psi$ suppression in
central nucleus-nucleus collisions at RHIC energies using a
multiphase transport model (AMPT) \cite{ampt1}. Since the AMPT model
includes both initial partonic and final hadronic interactions as well as 
the transition between these two phases of matter, it allows us to 
study the relative importance of the partonic and hadronic 
effects on $J/\psi$ suppression.

This paper is organized as follows. In Section \ref{ampt}, we briefly
describe the AMPT model for ultrarelativistic heavy ion collisions
and its extension to include $J/\psi$ production. The mechanisms for
$J/\psi$ suppression are given in
Section \ref{absorption}. Results for Au+Au and S+S collisions
at RHIC are given in Section \ref{results}. First, we show
the time evolution of the particle number and energy densities in both
the partonic and hadronic matters. This is then followed by 
the time evolution of various $J/\psi$ suppression effects and the 
$J/\psi$ survival probability. Finally, a summary is given in 
Section \ref{summary}.   
 
\section{multiphase transport model}\label{ampt}

In the AMPT model, the initial conditions are obtained from  
the minijets generated by the Heavy Ion Jet Interaction Generator (HIJING) 
\cite{hijing1} by using a Woods-Saxon radial shape for the   
colliding nuclei and including the nuclear shadowing effect on partons
via the gluon recombination mechanism by Mueller-Qiu \cite{amueller1}. 
After the passage of the 
colliding nuclei, the Gyulassy-Wang model \cite{Gyulassy:1994hr} is then
used to generate the initial space-time information of partons.
The subsequent time evolution of the parton phase-space distribution is
modeled by the Zhang's Parton Cascade (ZPC) \cite{zpc1,Gyulassy:1997ib},
which at present includes only the gluon elastic scattering.
After partons stop interacting, the Lund string fragmentation model
\cite{sjostrand1,bandersson1} is used to 
convert them into hadrons. The dynamics of resulting hadronic matter
is described by a relativistic transport model
(ART) \cite{art1}. Details of the AMPT model can be found in \cite{ampt1}.
 
To include $J/\psi$ in the AMPT model, we use the perturbative approach 
\cite{wcassing1} as its production probability is small in heavy ion 
collisions. Specifically, the reaction $gg\rightarrow J/\psi g$ is selected 
from the PYTHIA program whenever there is an inelastic scattering between 
projectile and target nucleons. Instead of allowing 
the produced $J/\psi$ to undergo multiple interactions, fragmentation, 
and decay in the PYTHIA, its momentum is stored in a file and read into
HIJING. The transverse position of the $J/\psi$
is determined by propagating it to the time of 
formation from the average transverse position of the colliding
projectile and target nucleons.  Since 
the $J/\psi$ production probability is increased from 
$\sigma(NN\rightarrow J/\psi X)$ to $\sigma_{inel}$, each $J/\psi$ 
is given a probability of $\sigma(NN\rightarrow J/\psi X)/\sigma_{inel}$
in order to have the correct number in an event, given by 
the total probability of all $J/\psi$'s.
The $J/\psi$ formation time is taken to be 0.5 fm/$c$, which is
suggested by the virtuality argument of 
Kharzeev and Satz \cite{Kharzeev:1996br} and
is also consistent with the uncertainty principle used for estimating
the lifetime of an expanding $c\bar{c}$ \cite{Gerland:1998bz,Arleo:2000af}.  
Before this time, the $c\bar c$ pair is considered a precursor $J/\psi$.

\section{$J/\psi$ absorption}\label{absorption}

After a pair of $c\bar c$ is produced in the initial collisions, 
whether it can materialize as a $J/\psi$ depends on its interactions
in the initial partonic and final hadronic matters. In the partonic matter, 
the $c\bar c$ pair, which are initially close in phase space and would
normally form a $J/\psi$ after the formation time of 0.5 fm/$c$, will 
move beyond the confinement distance of $J/\psi$ and become unbound
if the plasma screening remains strong \cite{Matsui:1986dk}. They will later 
combine with the more abundant light quarks to form instead the
charm hadrons when the parton density is low.  The critical
density for the plasma screening to be effective can be estimated
using $n_c=(2k_0 \mu_c^2)/(3\pi^2\alpha_S)$ \cite{Biro:1992ix},
where $\mu_c$ is the critical Debye screening mass, $\alpha_S$
is the QCD coupling constant, and $k_0$ is the slope parameter
of the transverse momentum distribution. Using
$\mu_c=0.7$ GeV, $\alpha_S=0.47$, and $k_0=0.6$ GeV, 
one obtains a critical density of $~5$ fm$^{-3}$. 
Following the method of Ref. \cite{Satz:2000wv}, we first determine 
from the AMPT model the time evolution of the radius
of the volume in which the parton density is above the critical density.  
A $J/\psi$ is thus not formed from the $c\bar c$ pair if its radial position
after the formation time of 0.5 fm/$c$ is within the critical radius
at that time.

Besides dissociation due to plasma screening, a $J/\psi$ 
may also be dissociated by collisions with gluons
in the partonic matter \cite{eshuryak1}. The cross
section for $gJ/\psi\rightarrow c\bar{c}$ has been estimated
in Ref. \cite{eshuryak1} to be 3 mb. Similarly, a precursor $J/\psi$ 
can also be dissociated by gluon scattering. The cross section is
expected to be smaller as its size is smaller than 
a physical $J/\psi$. In the present study, we ignore the
difference between the scattering of a gluon with precursor and physical
$J/\psi$, and use the same cross section. Since this cross section is
not well-known, we have also used a value of $1$ mb in the following study.

In the hadronic matter following the parton stage, 
a $J/\psi$ can be further destroyed by collisions with hadrons. The $J/\psi$
absorption cross section by baryons is taken to be 6 mb \cite{hufner} 
while that by meson is taken to be 3 mb \cite{gavin} above their 
respective thresholds. 
Although earlier studies based on the perturbative QCD \cite{satz} and 
simple hadronic model \cite{muller}
give a much smaller $J/\psi$ absorption cross sections by hadrons, 
the above values are consistent with recent studies using the 
quark-interchange model \cite{wong} and the more complete hadronic model 
\cite{haglin,lin}. 
Using these cross sections in the transport model, it has been found that
the observed suppression of $J/\psi$ production in heavy ion collisions
at the SPS energies can be reasonably described except for the most central
collisions \cite{cmk,Cassing:1999es,bhsa}. 
As the formation time of $J/\psi$ is comparable to the passing time
($0.1$ fm/$c$) of two colliding nuclei at RHIC energies, $J/\psi$ absorption
by the projectile and target nucleons is not expected to be important 
\cite{Hufner:1998hf} and will be ignored in the present study.
We note that as the $J/\psi$ is treated perturbatively in the simulation, 
the momenta of other hadrons are not affected by their interactions with 
$J/\psi$ \cite{Fang:1994cm}.

\section{Results}\label{results}

\subsection{time evolution of particle number and energy densities}

\begin{figure}
\centerline{\epsfig{file=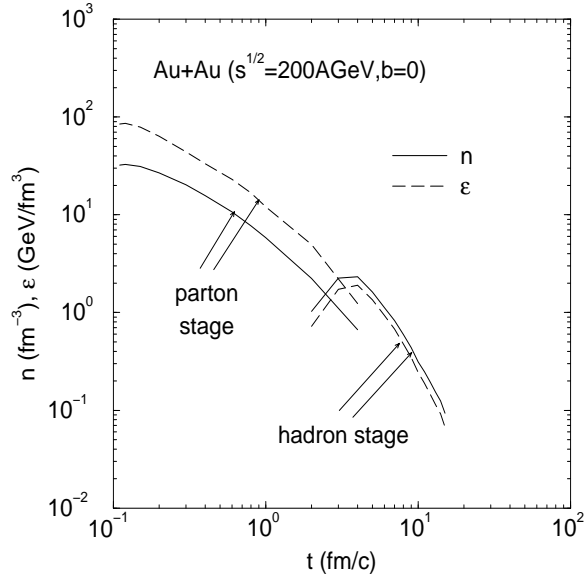,height=3.0in,width=3.0in,angle=-90}}
\vspace{0.3in}
\caption{
Time evolution of particle number and energy densities 
in the central cell around $\eta=0$ and $r=0$
for central (b=0) Au+Au collisions at RHIC. 
\label{fig:auauden5}}
\end{figure}

\begin{figure}
\centerline{\epsfig{file=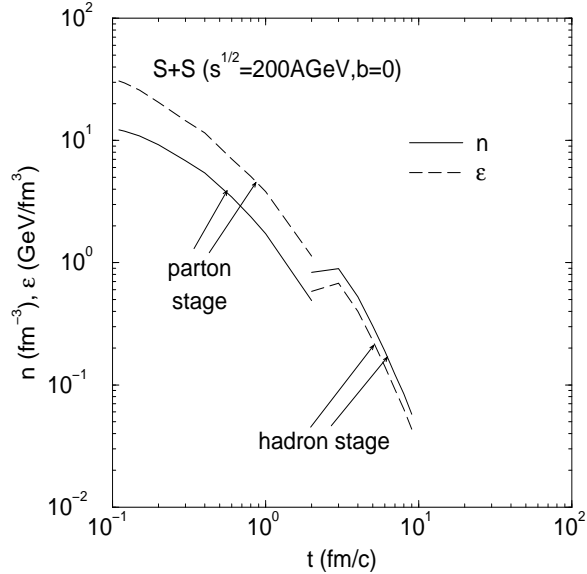,height=3.0in,width=3.0in,angle=-90}}
\vspace{0.3in}
\caption{
Same as Fig. \ref{fig:auauden5} for central (b=0) S+S collisions at RHIC. 
\label{fig:ssden9}}
\end{figure}

In Fig. \ref{fig:auauden5}, we show the time evolution of the 
particle number and energy densities of partons and hadrons 
in central Au+Au collisions at $\sqrt{s}=200$ GeV. They are
the densities in the central cell, which has a transverse radius of 1 fm 
for partons and 2 fm for hadrons. The longitudinal dimension of 
the central cell is taken to be $\pm 5\%$ of the time $t$, which is 
equivalent to taking the central space-time rapidity cell. The results
are not significantly changed if $\pm 10\%$ of the time is used 
for the longitudinal dimension. We 
see that the initial parton density is about $30$ fm$^{-3}$ and 
is much higher than the $J/\psi$ dissociation critical density 
(about $5$ fm$^{-3}$) given above. The time evolution of these
densities is seen to deviate from that 
based on the ideal Bjorken boost invariant scenario \cite{Bjorken:1983qr}, 
which would lead to a linear curve on the log-log plot.
This difference is mainly due to the more realistic 
treatment of initial collisions by using the
the Gyulassy-Wang model for the formation time and
the presence of radial flow as well as a
gradual freeze-out \cite{dmolnar} 
at the later hadronic stage. From the ratio of the 
parton energy density to its number density, one sees that the average
energy of a parton is more than $1$ GeV. The parton stage lasts about
$2-3$ fm/$c$, while the hadron stage starts gradually at around $3-4$ fm/$c$
and lasts until about $10$ fm/$c$. The average energy of a hadron
is less than $1$ GeV, since the central rapidity is
dominated by mesons instead of baryons in ultrarelativistic nuclear 
collisions. In Fig. \ref{fig:ssden9}, we show 
the results for central collisions of S+S at $\sqrt{s}=200$ AGeV. Due to
the smaller size of the system comparing to that of  
Au+Au collisions, the plasma lifetime is shorter, i.e., about $1-2$ fm/$c$,
and the hadron stage sets in at about $2-3$ fm/$c$.
We note that even in the smaller parton system produced in S+S collisions, 
the initial density is much higher than the critical density
for $J/\psi$ dissociation.

\begin{figure}
\centerline{\epsfig{file=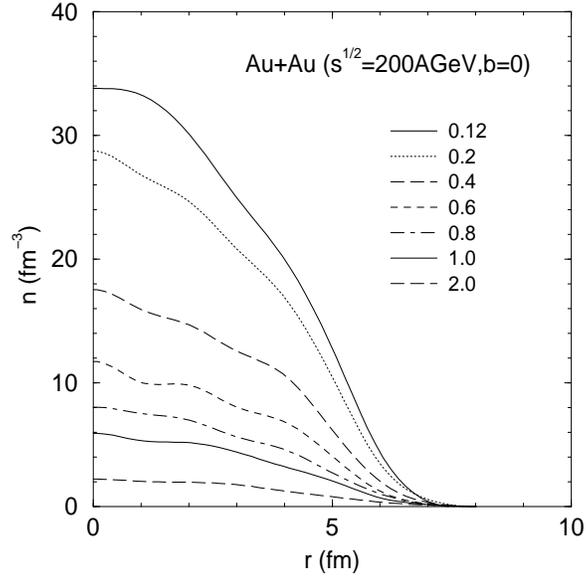,height=3.0in,width=3.0in,angle=-90}}
\vspace{0.3in}
\caption{
Parton density in central (b=0) Au+Au collisions at RHIC
at different times. The numbers are the proper times in fm/$c$ for
the density measurements.
\label{fig:auaurho4c}}
\end{figure}
 
\begin{figure}
\centerline{\epsfig{file=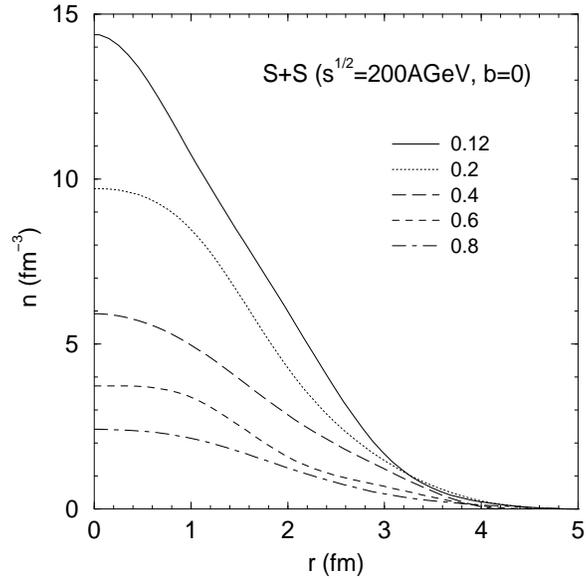,height=3.0in,width=3.0in,angle=-90}}
\vspace{0.3in}
\caption{
Parton density in central (b=0) S+S collisions at RHIC at different times.
\label{fig:ssrho4a6}}
\end{figure}
 
\begin{figure}
\centerline{\epsfig{file=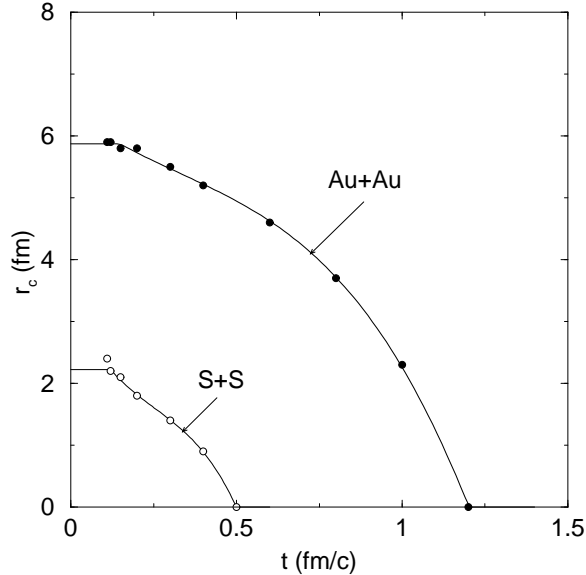,height=3.0in,width=3.0in,angle=-90}}
\vspace{0.3in}
\caption{
Time evolution of the critical radius for plasma dissociation of 
$J/\psi$ in central Au+Au and S+S collisions at $\sqrt{s}=200$ AGeV.
\label{fig:auaurhoc3}}
\end{figure}

In Fig. \ref{fig:auaurho4c} and Fig. \ref{fig:ssrho4a6}, we
show, respectively, the parton density in central Au+Au and S+S 
collisions at different times. As minijet gluons are produced from 
initial hard collisions, their densities
at different radii reflect the number of initial
binary nucleon-nucleon collisions. As seen from the figure, 
they are different from 
the initial nuclear density distribution given by a Woods-Saxon form. 
As expected,  both
the size and lifetime of the partonic matter produced in S+S collisions 
are smaller than those in Au+Au collisions.
From the time evolution of the parton density,  
one can determine the time evolution of the critical radius for
$J/\psi$ dissociation, and this is shown in Fig. \ref{fig:auaurhoc3}  
by full dots for Au+Au collisions and open dots
for S+S collisions. The solid lines are 
polynomial fits to the above results using the form, 
$r_c(t)=a_0+a_1t+a_2t^2+a_3t^3$ for $t_{min}<t<t_{max}$, with 
$r_c(t)=r_c(t_{min})$ for $t<t_{min}$ and  $r_c(t)=0$ for $t>t_{max}$.
For the partonic matter produced in Au+Au collisions, we have 
$t_{min}=0.15$ fm/$c$, $t_{max}=1.2$ fm/$c$, $a_0=6.39$, $a_1=-4.06$,
$a_2=4.79$, and $a_3=-4.86$. For S+S collisions, they are 
$t_{min}=0.12$ fm/$c$,
$t_{max}=0.5$ fm/$c$, $a_0=3.278$, $a_1=-11.949$, $a_2=31.514$, 
and $a_3=-41.428$. In Au+Au collisions,  
the critical radius for $J/\psi$ dissociation extracted from our model is  
about 6 fm at beginning of the parton cascade and vanishes after about
1.2 fm/$c$. The critical radius is reduced to about 2 fm initially and 
vanishes after only about 0.5 fm in S+S collisions.   Since
the duration of the partonic matter that is above the critical density 
for $J/\psi$ dissociation is shorter than the $J/\psi$ formation time, 
$J/\psi$ suppression due to plasma screening is therefore unimportant 
in central S+S collisions.

\subsection{time evolution of $J/\psi$ survival probability}

\begin{figure}
\centerline{\epsfig{file=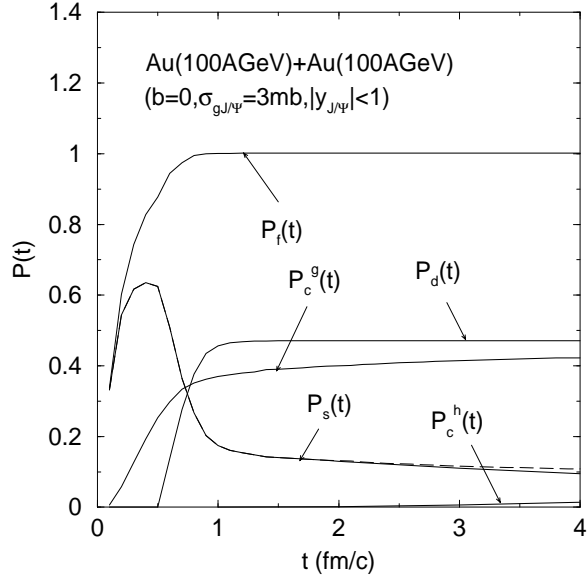,height=3.0in,width=3.0in,angle=-90}}
\vspace{0.3in}
\caption{
Time evolution of the $J/\psi$ formation probability $P_f(t)$,
the $J/\psi$ absorption probability $P_c^g(t)$
by gluon scattering, the $J/\psi$ dissociation probability $P_d(t)$
by plasma screening, the $J/\psi$ absorption probability $P_c^h(t)$ 
by hadrons, and the $J/\psi$ survival probability $P_s(t)$. 
For the survival probability, the solid curve includes absorption 
by both gluons and hadrons, while 
the long-dashed curve includes only absorption by gluons.
\label{fig:prob2a}}
\end{figure}

Including the above three mechanisms for $J/\psi$ suppression in the
AMPT model, we have evaluated the $J/\psi$ survival probability
in ultrarelativistic heavy ion collisions at RHIC energies.
In Fig. \ref{fig:prob2a}, we show for  central Au+Au 
collisions the time evolution of the $J/\psi$ formation
probability $P_f(t)$, the $J/\psi$ absorption probability $P_c^g(t)$
by gluon scattering, the $J/\psi$ dissociation probability $P_d(t)$
by plasma screening, the $J/\psi$ absorption probability $P_c^h(t)$ 
by hadrons, and the $J/\psi$ survival probability 
$P_s(t)=1-P_c^g(t)-P_d(t)-P_c^h(t)$. The results are taken 
for the $J/\psi$ produced in
the central rapidity interval $|y_{J/\psi}|<1$. It is seen
that the $J/\psi$ formation probability
increases quickly with time and about $90\%$ of
the $J/\psi$ are formed by $0.5$ fm/$c$. 
Absorption by gluons starts very early in the
process and ends at about $1$ fm/$c$ after dissociation due to 
plasma screening begins. The latter also ends at about $1$ fm/$c$. 
Both dissociation due to plasma screening and absorption by gluon
collisions give comparable contributions, accounting for the 
suppression of about 90\% of the $J/\psi$, 
while absorption by hadrons contributes only about 
a few percent. The final $J/\psi$ survivability is about 6\%.

\begin{figure}
\centerline{\epsfig{file=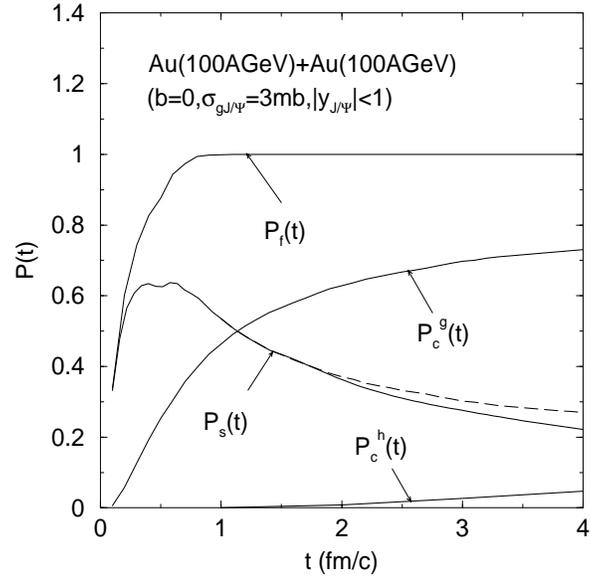,height=3.0in,width=3.0in,angle=-90}}
\vspace{0.3in}
\caption{
Same as \ref{fig:prob2a} without plasma dissociation.
\label{fig:prob2b}}
\end{figure}

\begin{figure}
\centerline{\epsfig{file=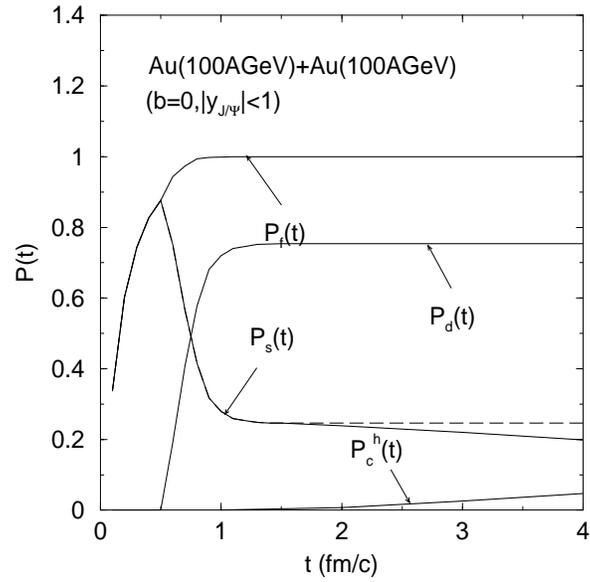,height=3.0in,width=3.0in,angle=-90}}
\vspace{0.3in}
\caption{
Same as \ref{fig:prob2a} without gluon destruction.
\label{fig:prob2c}}
\end{figure}

\begin{figure}
\centerline{\epsfig{file=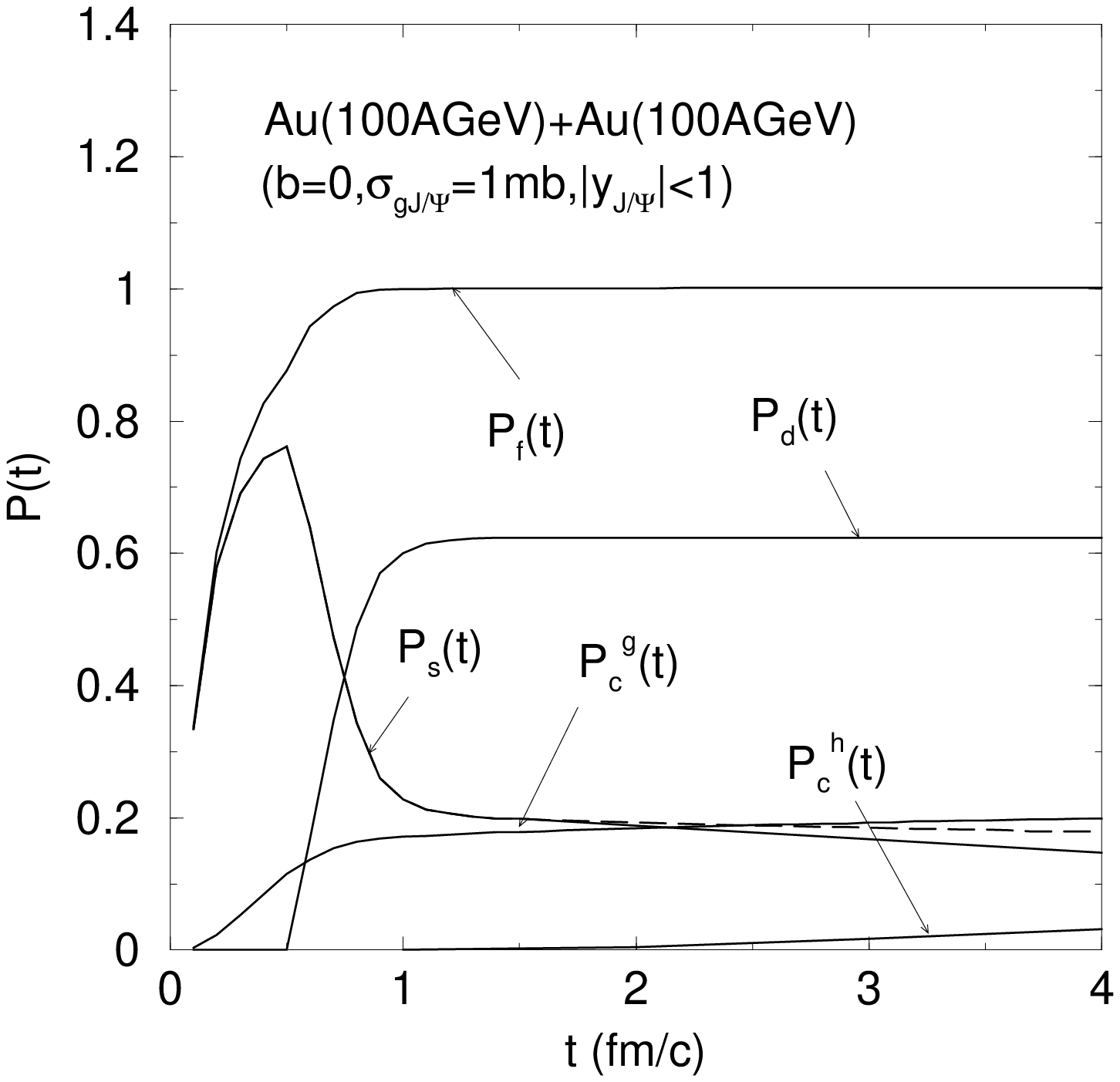,height=3.0in,width=3.0in,angle=-90}}
\vspace{0.3in}
\caption{
Same as \ref{fig:prob2a} with reduced $J/\psi$ absorption cross section
by gluons.
\label{fig:prob2d}}
\end{figure}

The results for the case without plasma screening is shown in
Fig. \ref{fig:prob2b}. Comparing to Fig. \ref{fig:prob2a},
we see that the $J/\psi$ absorption probability by gluons 
is similar for time up to $0.5$ fm/$c$, but it lasts longer 
until $2-3$ fm/$c$.  This indicates that there is a competition between 
$J/\psi$ suppression due to gluon scattering and 
plasma screening, i.e., some $J/\psi$'s that are destroyed by 
collisions with gluons would have been dissociated
by plasma screening if they are not allowed to scatter with gluons.
Fig. \ref{fig:prob2c} shows the results without $J/\psi$ absorption
by gluon scattering.
In this case, the plasma dissociation
probability saturates quickly. Compared to that shown in 
Fig. \ref{fig:prob2b}, the $J/\psi$ suppression probability is seen to  
start late but saturate early. 
This difference in the survival probability may be seen by 
studying the azimuthal distribution of final survival $J/\psi$ 
in mid-central collisions. It is interesting 
to note that $J/\psi$ suppression during the parton stage is
similar whether when both plasma screening and gluon scattering
are present or when only one of them is present.
The effect of reducing the $J/\psi$ scattering cross section by gluon 
is shown in
Fig. \ref{fig:prob2d}. Compared with Fig. \ref{fig:prob2a}, we see
that the decrease in the contribution from 
gluon scattering is partly compensated by the plasma screening.

In Table \ref{table:auauprob}, we summarize 
the final absorption probabilities due to gluons ($P_c^g$), hadrons
($P_c^h$), plasma dissociation ($P_d$), and the survival probability 
($P_s$) in central Au+Au collisions. 
It is seen that the $J/\psi$ survival probability
after the parton stage, $S_g=1-P_c-P_d$, is
about 9\% and 17\% for $\sigma_{gJ/\psi}=$ 3 mb and 1 mb, respectively.
The parton stage thus has a large 
effect on $J/\psi$ suppression in heavy ion collisions at RHIC.
In both cases, the $J/\psi$ suppression factor in the hadron stage is
$S_h=P_h/S_g\sim 35\%$. 
Depending on the different assumptions on the mechanisms for
$J/\psi$ suppression, the final survival probability can range
from about $6\%$ to about $26\%$. This indicates that finite size effects 
may be important in $J/\psi$ suppression at RHIC energies.
We have also studied $J/\psi$ suppression in S+S collisions at RHIC energies,
and the results are summarized in Table \ref{table:ssprob}.
As expected from the discussion of Fig. \ref{fig:auaurhoc3}, there is
no contribution from plasma screening in collisions of such light system,
although the initial density is above the critical density.  
As a result, the $J/\psi$ has an appreciable
survival probability (about 50\%) after absorption by gluons and hadrons.
Since $J/\psi$ absorption by gluon scattering still contributes 
appreciably to the final $J/\psi$ suppression, it may provide an 
opportunity for studying this mechanism.
We note that the $J/\psi$ suppression factor in the hadron stage is about
$S_h\sim 16\%$, which is smaller than in Au+Au collisions.

\begin{table}
\caption{Final 
$J/\psi$ absorption probability $P_c^g$
by gluon scattering, $J/\psi$ dissociation probability $P_d$ by plasma 
screening, $J/\psi$ absorption probability $P_c^h$ by hadrons, 
and $J/\psi$ survival probability $P_s$
for Au+Au ($b=0$, $\sqrt{s}=200$ AGeV).
$(gJ/\psi)_a$ indicates $\sigma_{gJ/\psi}=3$ mb while $(gJ/\psi)_b$ 
indicates $\sigma_{gJ/\psi}=1$ mb.
}
\label{table:auauprob}
\begin{tabular}{lcccc}
 & $P_c^g$ & $P_d$ & $P_c^h$ & $P_s$ \\ \tableline
screening+$(gJ/\psi)_a$+$hJ/\psi$ 
& $43.9\%$ & $47.1\%$ & $ 2.9\%$ & $ 6.1\%$ \\
screening+$(gJ/\psi)_b$+$hJ/\psi$
& $20.6\%$ & $62.4\%$ & $ 6.5\%$ & $10.5\%$ \\
screening+h$J/\psi$
& $0$      & $75.4\%$ & $10.2\%$ & $14.4\%$ \\
$(gJ/\psi)_a$+$hJ/\psi$ 
& $78.3\%$ & $0$      & $10.6\%$ & $11.0\%$ \\
$(gJ/\psi)_b$+$hJ/\psi$ 
& $42.9\%$ & $0$      & $31.1\%$ & $26.0\%$ \\
$hJ/\psi$
& $0$      & $0$      & $56.7\%$ & $43.3\%$ \\
\end{tabular}
\end{table}

\begin{table}
\caption{Same as Table \ref{table:auauprob}
for S+S ($b=0$,$\sqrt{s}=200$ AGeV).
}
\label{table:ssprob}
\begin{tabular}{lcccc}
 & $P_c^g$ & $P_d$ & $P_c^h$ & $P_s$ \\ \tableline
screening+$(gJ/\psi)_a$+$hJ/\psi$ 
& $39.3\%$ & $0$      & $10.6\%$ & $50.1\%$ \\
screening+$(gJ/\psi)_b$+$hJ/\psi$ 
& $16.8\%$ & $0$      & $16.0\%$ & $67.2\%$ \\
$hJ/\psi$
& $0$      & $0$      & $19.6\%$ & $80.4\%$ \\
\end{tabular}
\end{table}

\section{summary}\label{summary}

In summary, we have studied $J/\psi$ suppression in Au+Au and S+S collisions
at RHIC energies in a multiphase transport model. It is found
that finite size effects
may be important in $J/\psi$ suppression at RHIC energies even in 
central Au+Au collisions. Both $J/\psi$ suppression due to 
plasma screening and gluon scatterings are important in Au+Au collisions, 
leading to a final $J/\psi$ survival probability of  
only about $6\%$.
For S+S collisions, even though the initial energy density is above the
critical density for $J/\psi$ dissociation, 
only $J/\psi$ absorption by gluon and hadron scatterings 
contribute to its suppression as a result of the short lifetime
of the plasma. $J/\psi$ suppression in heavy ion collisions 
between light nuclei thus provides a possible opportunity for studying 
the interactions between $J/\psi$ and gluon.

In our study, we have ignored both $J/\psi$ formation from 
the recombination of $c\bar c$ pairs during hadronization
and $J/\psi$ production from charm mesons in the final hadronic matter.
The number of $J/\psi$ formed from charm quarks during hadronization
can be roughly estimated from considerations of the phase
space and the spin and color factors. At the end of parton stage, 
the volume of the partonic matter in the central unit of rapidity
is approximately given by $V=\pi R^2\tau$, where $R$ and $\tau$
are the transverse radius and longitudinal dimension, respectively.
For Au+Au collisions, they are $R\sim 7~{\rm fm}$ and $\tau\sim 3~{\rm fm}$.
The probability that an anticharm quark is found within a
distance of the $J/\psi$ radius from a charm quark is given by the ratio of 
the size of the $J/\psi$ to the size of the partonic matter.
With a $J/\psi$ radius $r_{J/\psi}\sim 0.3~{\rm  fm}$, one
obtains the probability $~2.5\times 10^{-4}$. In order for
this charm pair to form a $J/\psi$, their spins must be parallel and
their total color should be neutral. This reduces the probability by
the factor $(3/4)\cdot (1/9)\sim 0.083$. Requiring that the charm
pair has a relative momentum distribution within 
$p_0\sim 1/r_{J/\psi}\sim 0.67$ GeV at $T\sim 200$ MeV further
suppresss the probability by a factor of 0.29. Since there are about
1.5 charm pairs per unit rapidity per central Au+Au collision at RHIC,
the number of $J/\psi$ formed from their recombination is about
$1.5^2\cdot 2.5\times 10^{-4}\cdot 0.083\cdot 0.29\sim 1.4\times 10^{-5}$, 
which is much smaller than the expected number of $J/\psi$, i.e.,
$1.5/40\cdot 0.061\sim 2.3\times 10^{-3}$. In the above, the factor 40 is
the ratio of charm to $J/\psi$ production in initial hard collisions 
\cite{rvogt99} and
the factor 0.061 is the final $J/\psi$ survival probability from 
Table \ref{table:auauprob}.
As to $J/\psi$ production from charm mesons in hadronic matter, it 
has been shown to be insignificant in
heavy ion collisions at RHIC, although it may not be negligible at LHC
\cite{ko,redlich}. 

We have not included the effect due to 
$\psi'$ and $\chi_c$, which are also produced in initial collisions
and can decay to $J/\psi$. Since these particles are less bound than
$J/\psi$, they are more likely to be dissociated and absorbed
in both the quark-gluon plasma and the hadronic matter. As a result,
inclusion of $\psi^\prime$ and $\chi_c$ is expected to lead to a smaller 
number of $J/\psi$ than predicted in the present study.
Also, we have used the default gluon shadowing in HIJING without 
flavor and scale dependence,
which is different from the one suggested by Eskola \cite{eskola}
that includes such dependence. 
If the latter is used, then we expect a reduction of the $J/\psi$
survival probability as the gluon density would be higher at RHIC energies
\cite{xnwang}. These effects will be studied in the future.

\section{Acknowledgments}

The authors thank M. Gyulassy and J. Nagle for helpful discussions.
They also thank the Parallel Distributed System Facility at National
Energy Research Scientific Computer Center for providing computing
resources.
This work was supported by the National Science Foundation under Grant 
No. PHY-9870038, the Welch Foundation under Grant No. A-1358, and the Texas 
Advanced Research Program under Grant Nos. FY97-010366-0068 and
FY99-010366-0081. The work of B.A. Li was supported in part by the
NSF under Grant No. 0088934 and the Arkansas Science and Technology Authority 
under Grant No. 00-B-14. B.H. Sa was partially 
supported by the Natural Science Foundation and Nuclear Industry
Foundation of China.


\end{document}